\documentclass[review]{elsarticle}
\usepackage{amsmath}
\usepackage{lineno,hyperref}
\usepackage{booktabs}
\modulolinenumbers[5]

\journal{Phys. Lett. B}

\begin{document}

\begin{frontmatter}

\title{Deflated GMRES with Multigrid for Lattice QCD}
%\tnotetext[mytitlenote]{Fully documented templates are available in the elsarticle package on \href{http://www.ctan.org/tex-archive/macros/latex/contrib/elsarticle}{CTAN}.}

%% Group authors per affiliation:
\author[Bay1]{Travis Whyte}
\ead{travis\_whyte@baylor.edu}
%\email{travis\_whyte@baylor.edu}
%\affiliation{Department of Physics, Baylor University \\
%Waco, TX 76798-7316}
\author[Bay1]{Walter Wilcox}
\ead{walter\_wilcox@baylor.edu}
%\email{walter\_wilcox@baylor.edu}
%\affiliation{Department of Physics, Baylor University \\
%Waco, TX 76798-7316}
\author[Bay2]{Ronald B. Morgan}
\ead{ronald\_morgan@baylor.edu}
\address[Bay1]{Department of Physics, Baylor University, Waco, TX 76798-7316}
\address[Bay2]{Department of Mathematics, Baylor University, Waco, TX 76798-7316}

\begin{abstract}
Lattice QCD solvers encounter critical slowing down for fine lattice spacings and small quark mass. Traditional matrix eigenvalue deflation is one approach to mitigating this problem. However, to improve scaling we study the effects of deflating on the coarse grid in a hierarchy of three grids for adaptive mutigrid applications of the two dimensional Schwinger model. We compare deflation at the fine and coarse levels with other non deflated methods. We find the inclusion of a partial solve on the intermediate grid allows for a low tolerance deflated solve on the coarse grid. We find very good scaling in lattice size near critical mass when we deflate at the coarse level using the GMRES-DR and GMRES-Proj algorithms.
% took out the word also
\end{abstract}


\end{frontmatter}

%\linenumbers

\section{Introduction}

The problem of critical slowing down and strong scaling is one of the foremost problems facing modern Lattice Quantum Chromodynamics (LQCD) simulations. Current simulations require extremely fine lattice spacings, necessitating the need for larger lattice volumes. This in turn creates larger Dirac operators needed for linear systems calculations such as stochastic trace estimators and fermionic forces in Hybrid Monte Carlo. Moreover, as the fermion mass approaches physically relevant values, the Dirac operator becomes extremely ill conditioned. This ill conditioning leads to exceptional eigenvalues, which drastically slow convergence of linear equations. Adaptive multigrid (MG)\cite{Brezina} is one method that deals with both the strong scaling and critical slowing down at the same time, and has been used successfully for the Wilson, overlap and staggered fermion discretizations \cite{Babich:2010qb,Brannick:2014vda,Brower:2018ymy}. Adaptive MG creates a hierarchy of coarsened operators from the original fine Dirac operator by exploiting its near null kernel. This shifts critical slowing down to the coarsest level, where the components of the error attributed to the exceptional eigenvalues can be more easily dealt with. However, the cost of the coarse grid solve can be very large when cast in terms of fine grid equivalence. 

Deflation has long been used as a method of dealing with exceptional eigenvalues in many fields, but is not yet heavily used in modern LQCD simulations, partly because of eigenvector storage costs for large systems. Adaptive MG allows for deflation to be employed on the coarsest level, where storage requirements of deflation are much smaller\cite{HISQgpu,Clark:2017wom,Yang}. The preferred method of MG in LQCD is to use it as a preconditioner for an outer Krylov solver \cite{Brannick:2007cc}. Because every iteration of the outer Krylov solver represents a new right hand side for the MG preconditioner, deflation with projection methods\cite{GMRES-DR,gproj} can be efficiently employed on the coarsest level. We demonstrate the effect that deflation on the coarsest level has by comparing to MG without coarse grid deflation, and the effect that this deflation has for multiple right hand sides. We observe that a partial solve on the intermediate grid in conjunction with deflation and projection methods on the coarse grid allows for a partial coarse grid solve. This partial solve on the intermediate grid reduces the number of outer iterations for convergence, and we observe no sign of critical slowing down resurgence on the higher grid levels with the use of a deflated partial coarse grid solve.
% Removed an extra it is observed

\section{Methods}

We work with the Wilson-Dirac operator in the two-dimensional lattice Schwinger Model\cite{Schwinger:1962tp}, which shares many physical characteristics with 4D LQCD, and as such is a good algorithmic testing ground. We created 10 gauge configurations within QCDLAB 1.0\cite{Borici:2006ch} for lattices of size $64^2$, $128^2$ and $256^2$ at $\beta = 6.0$. All values are averaged over seperate solves for each configuration.The method of coarsening follows that of reference \cite{Brower:2018ymy}. A hierarchy of three grids was created by solving the residual system $DD^{\dagger}e = -DD^{\dagger}x$, where $x$ is a random vector, for 12 near null vectors on the fine grids. This system was solved to a tolerance of $10^{-4}$, and the near null vectors were constructed using $\psi = x + e$. The near null vectors are globally orthonormalized, then subsequently chirally doubled using the projectors $\frac{1}{2}(1 \pm \sigma_3)$. They are then locally blocked and locally orthonormalized using a $4^2$ grid within the lattice to form the columns of the prologantor matrix, $P$. The intermediate grid operator $\hat{D}$, is then formed via $\hat{D} = P^{\dagger}DP$. This process is repeated to form the coarse grid operator.

As an outer solver, we use FGMRES(8)\cite{Sa93}, and two iterations of GMRES\cite{SaSc} as a pre and post smoother on the fine and intermediate levels. For our deflated MG solve, we solve to a tolerance of $10^{-15}$ with GMRES-DR(200,40) on the coarse grid for the first outer iteration, and for each subsequent outer iteration use GMRES-Proj(200,40) to solve to a tolerance of $10^{-8}$. Performing the coarse level solve with GMRES-DR for the first outer iteration allows us to efficiently solve the linear equations and the eigenvalue problem at the same time. For comparison, we perform the same solve, but with CGNE on the coarsest level, where the linear equations were solved to a tolerance of $10^{-8}$ for every outer iteration of FGMRES. It was found that using restarted methods, such as GMRES(m) and BiCGStab(m), on the coarse level solve offered no improvement over CGNE in terms of matrix vector products. We also compare to the standard deflation methods GMRES-DR(200,40) and GMRES-Proj(200,40) for multiple right hand sides on the fine grid. It should be noted that the large restart length was used in order for the solves to converge in a reasonable time period since this is a difficult numerical problem. For consistency and comparison, we use the same restart length and deflation length on the coarse grid in our deflated coarse solve. Throughout this study, $Z(4)$ noise was used as a right hand side. 

A partial solve on the intermediate grid level is also considered, where the linear equations on the intermediate grid are solved to a tolerance of $.2\times{||b||}$. For our deflated solve on the coarse grid, we solve to a tolerance of $10^{-8}$ for the first outer iteration, followed by a projected solve to a tolerance of $10^{-2}$ for subsequent outer iterations. It was observed that relaxing the tolerance for the non deflated solve on the coarse grid in the same fashion resulted in a large increase in outer iterations of FGMRES, so a solve to a tolerance of $10^{-8}$ was performed. We remark that the inclusion of a W-cycle, where the coarse grid is visited twice for every outer iteration, as is performed in reference \cite{Babich:2010qb}, may have ameliorated this problem, albeit at the price of increased coarse and intermediate matrix vector products. Since we aim to reduce the overall cost of the full solve, this method was avoided. 

\section{Results}

An indication that critical slowing down has been relayed to the coarsest level is a constant number of fine operator applications for increasing lattice volume as the mass gap approaches zero.
\begin{figure}[!htpb]
\centering
\includegraphics[trim={2cm 7cm 2cm 9cm},clip,width=1.0\textwidth]{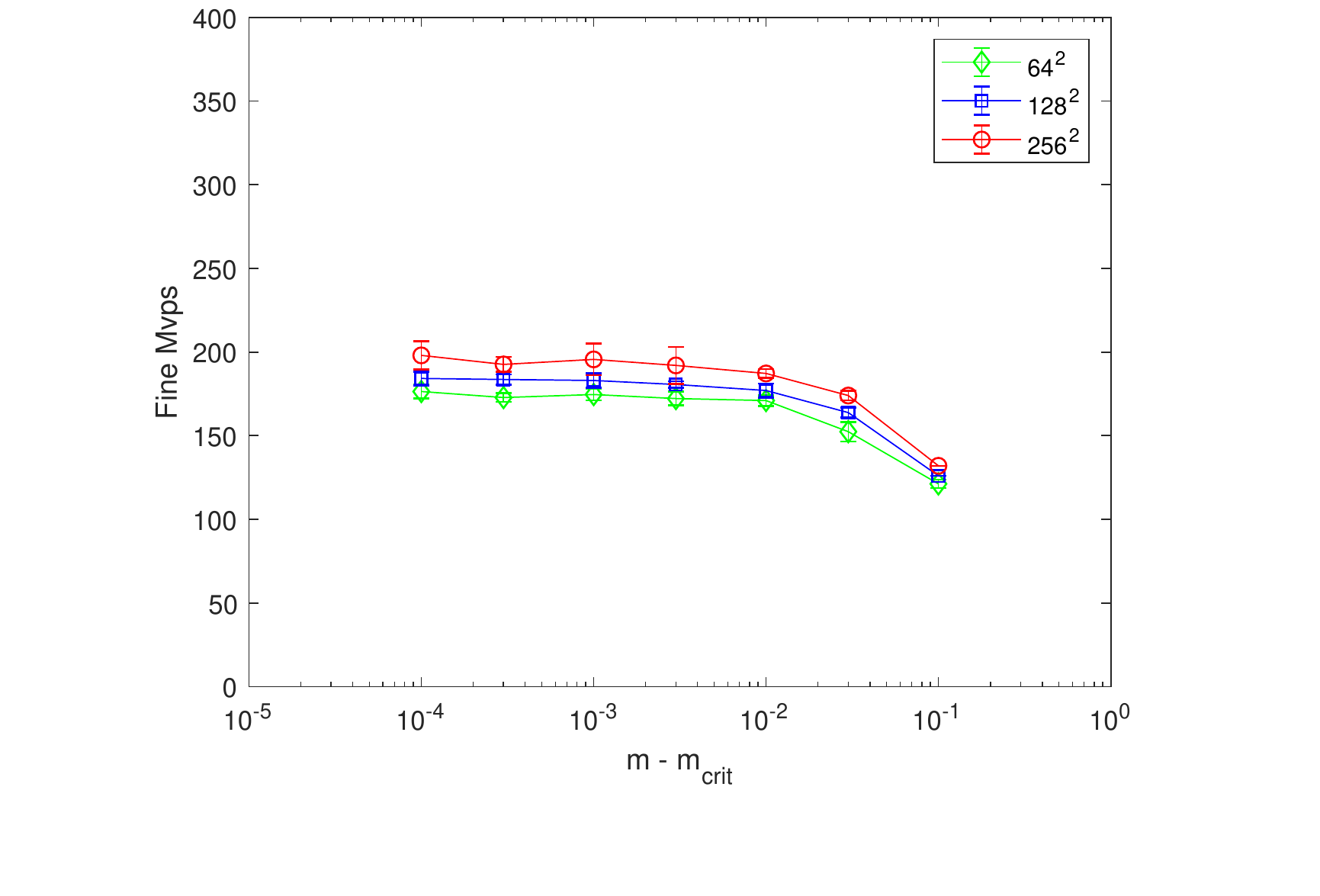}
\includegraphics[trim={2cm 8cm 2cm 9cm},clip,width=1.0\textwidth]{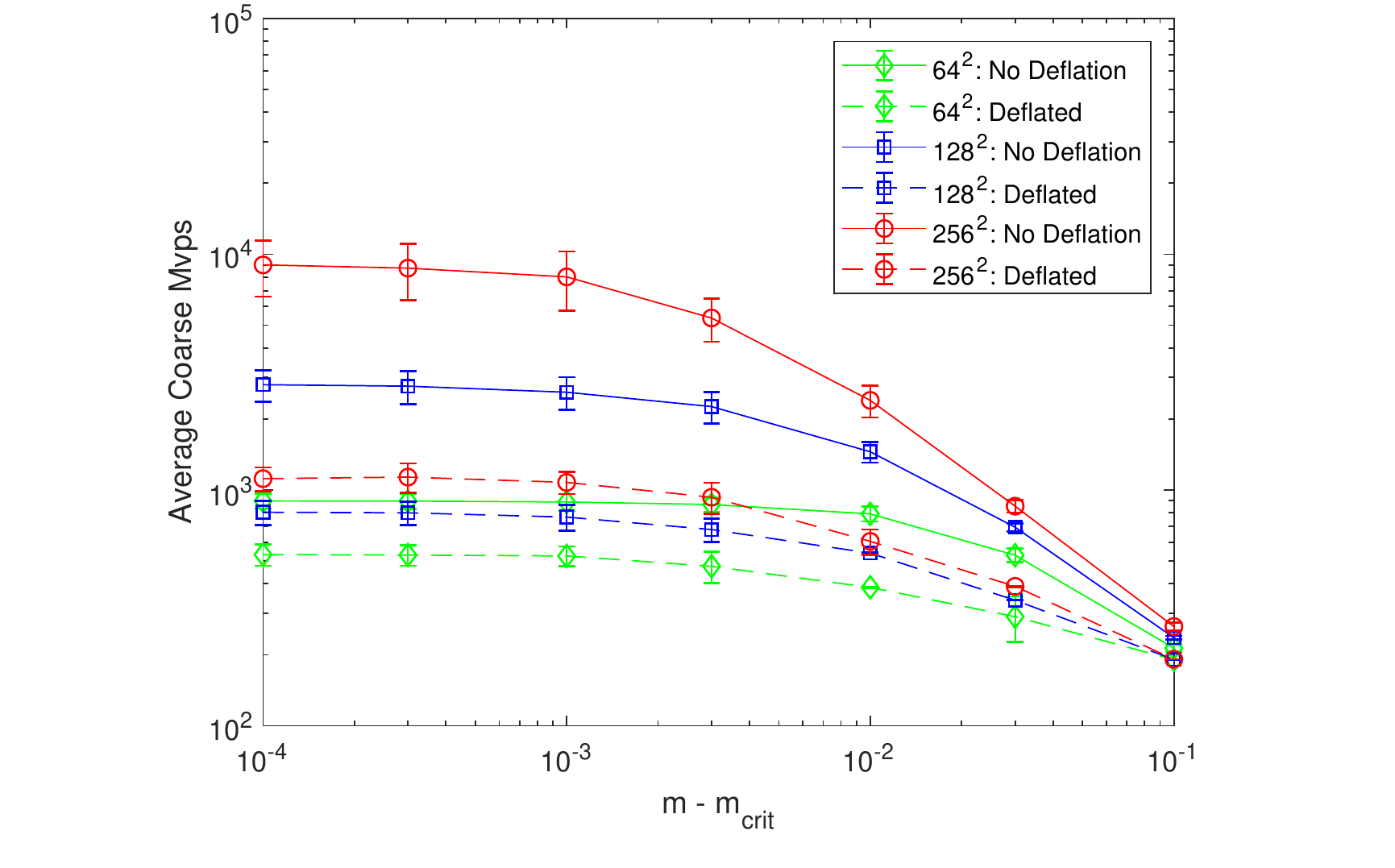}
\caption{(Top) The number of fine Dirac operator applications as a function of the mass gap for all three lattice sizes. (Bottom) The average number of coarse Dirac operator applications per outer iteration.}
\label{rad1}
\end{figure}
Figure 1 (Top) shows the number of fine Dirac operator applications for all three lattice sizes. The number of applications is nearly constant for all three lattice sizes as the mass gap approaches zero. Here, we only smooth on the intermediate level, so this is an indication that critical slowing down has been successfully shifted to the lowest level. Figure 1 (Bottom) shows the number of coarse operator applications averaged over the number of outer iterations as the mass gap approaches zero. The number of coarse operator applications increases drastically as the critical mass is approached, indicative of critical slowing down. Deflation significantly reduces the number of coarse operator applications. In the most dramatic case, our method of deflation has reduced the average number of coarse applications for the $256^2$ lattice to a number comparable to that of the $64^2$ lattice without deflation. We also observe the trend that as lattice size increases, deflation costs scale like a low power of lattice size. Since this is deflation with projection of eigenvectors on the coarsest level, this allows us to reap the benefits of deflation without the increased storage cost of retaining the fine eigenvectors and fine basis vectors. 
% Replaced drastically with significantly

We summarize the situation by recasting the coarse and intermediate grid operator applications in terms of fine equivalent matrix vector products. We define fine equivalent matrix vector products (Mvps) as:
\begin{equation}
\begin{aligned}
\textrm{Fine Equivalent Mvps}=&\\
\textrm{Fine Mvps} &+\displaystyle \frac{n_{int}}{n_{fine}} \times \textrm{Int Mvps} + \frac{n_{coarse}}{n_{fine}}\times \textrm{Coarse Mvps},
\end{aligned}
\end{equation}
where $n_{fine}, n_{int}$, and $n_{coarse}$ are the size of the Dirac operator for the fine, intermediate and coarse grids, respectively. Figure 2 shows a comparison of CGNE and GMRES-DR to non deflated and deflated MG preconditioned FGMRES. When evaluated in terms of fine equivalent Mvps, non deflated MG is nearly as expensive as CG on the normal equations. However, performing a deflated solve on the coarse grid drastically reduces the number of fine equivalent Mvps. It outperforms MG without deflation, and is more effective than pure deflation on the finest grid. 

\begin{figure}[!htpb]
\centering
\includegraphics[trim={2cm 9cm 2cm 9cm},clip,width=1.0\textwidth]{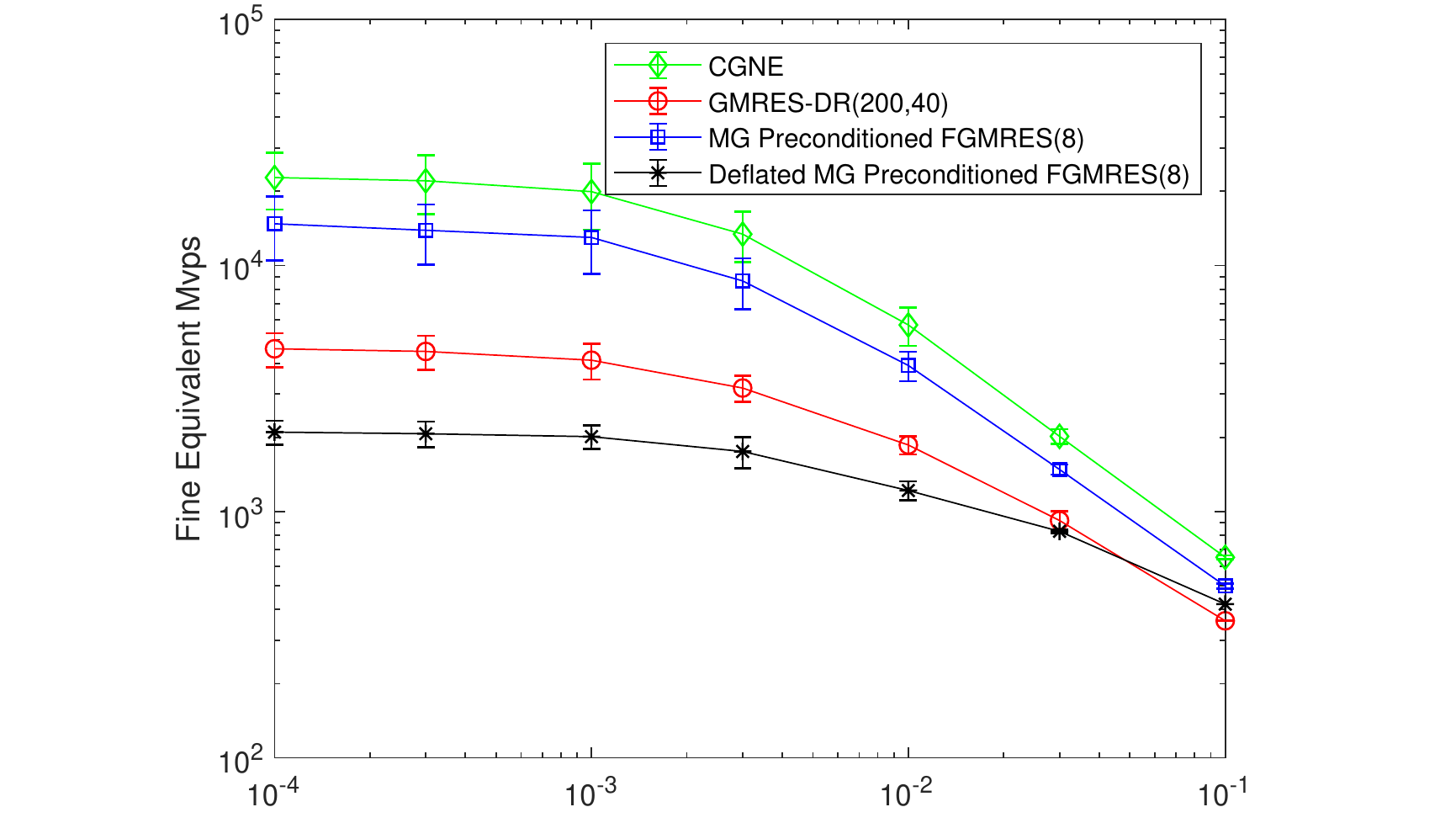}
\caption{A comparison of fine equivalent Mvps as a function of the mass gap for CGNE, GMRES-DR, MG and deflated MG for the lattice of size $256^2$.}
\label{rad2}
\includegraphics[trim={2cm 9cm 2cm 8cm},clip,width=1.0\textwidth]{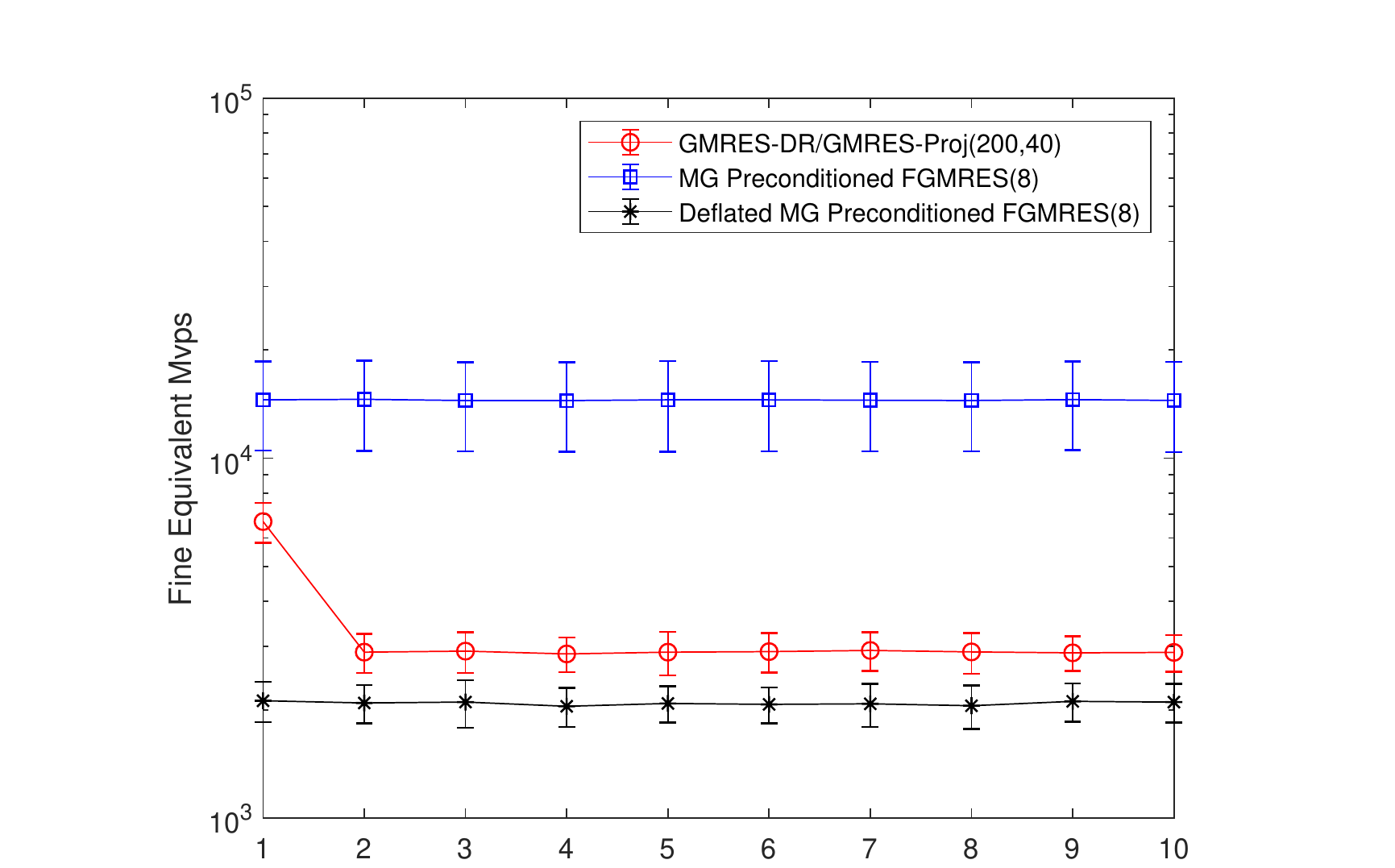}
\caption{A comparison of fine equivalent Mvps for multiple right hand sides for GMRES-DR/GMRES-Proj, MG and deflated MG at critical mass for the lattice of size $256^2$.}
\label{rad3}
\end{figure}
%\begin{figure}[ht]
%\centering
%\includegraphics[trim={2cm 8cm 2cm 8cm},clip,width=1.0\textwidth]{inverter_lattice_size_comparison_with_partial_solve_adjusted.pdf}
%\caption{Fine equivalent Mvps as a function of lattice size at a mass gap = $10^{-4}$.}
%\label{rad4}
%\end{figure}

We also observe that coarse grid deflation is more effective than traditional deflation on the fine grid for multiple right hand sides. Figure 3 displays the fine equivalent Mvps for a solve with ten right hand sides for the $256^2$ lattice at critical mass. The first right hand side was oversolved to a tolerance of $10^{-15}$ for our GMRES-DR/GMRES-Proj runs. This was done to obtain eigenvectors accurate enough to deflate with GMRES-Proj, the cost being amortized over subsequent right hand sides. On the coarse grid, only the first solve of the first outer iteration was oversolved to a tolerance of $10^{-15}$ to obtain both the solution to the linear equations and the eigenvectors on the coarse grid. Subsequent outer iterations and right hand sides were solved to a tolerance of $10^{-8}$ with GMRES-Proj. Once again, coarse deflation outperforms fine deflation and standard MG. We see that GMRES-DR computes the eigenvectors contributing to critical slowing down, while MG shifts critical slowing down to the coarse level where deflation then removes their contribution. We would n\"aively expect the results for our fine deflation solve and coarse deflation MG solve to be approximately equal over multiple right hand sides since both are dealing with the same deflated system. Instead, our results suggest that MG and coarse grid deflation act in a synergistic way, improving the effectiveness of the solve that neither method can achieve on their own.
% umlaut on naively added

We now briefly consider the effect of a partial solve on the intermediate grid. The inclusion of a partial solve on the intermediate level reduces the number of outer iterations for the full solve. Figure 4 displays the number of fine Dirac operator applications (Top) for both deflated and non deflated MG for the $256^2$ lattice, and the number of average intermediate operator applications (Bottom).  Both the deflated and non deflated methods display a reduction of fine Mvps compared to only smoothing on the intermediate level. We observe that the average number of intermediate operator applications remains constant as a function of mass gap, indicating that critical slowing down has not been shifted back up to the intermediate level, despite the low tolerance solve on the coarse grid for our deflated method.
%mvps to Mvps
\begin{figure}[!htpb]
\centering
\includegraphics[trim={2cm 7cm 2cm 9cm},clip,width=1.0\textwidth]{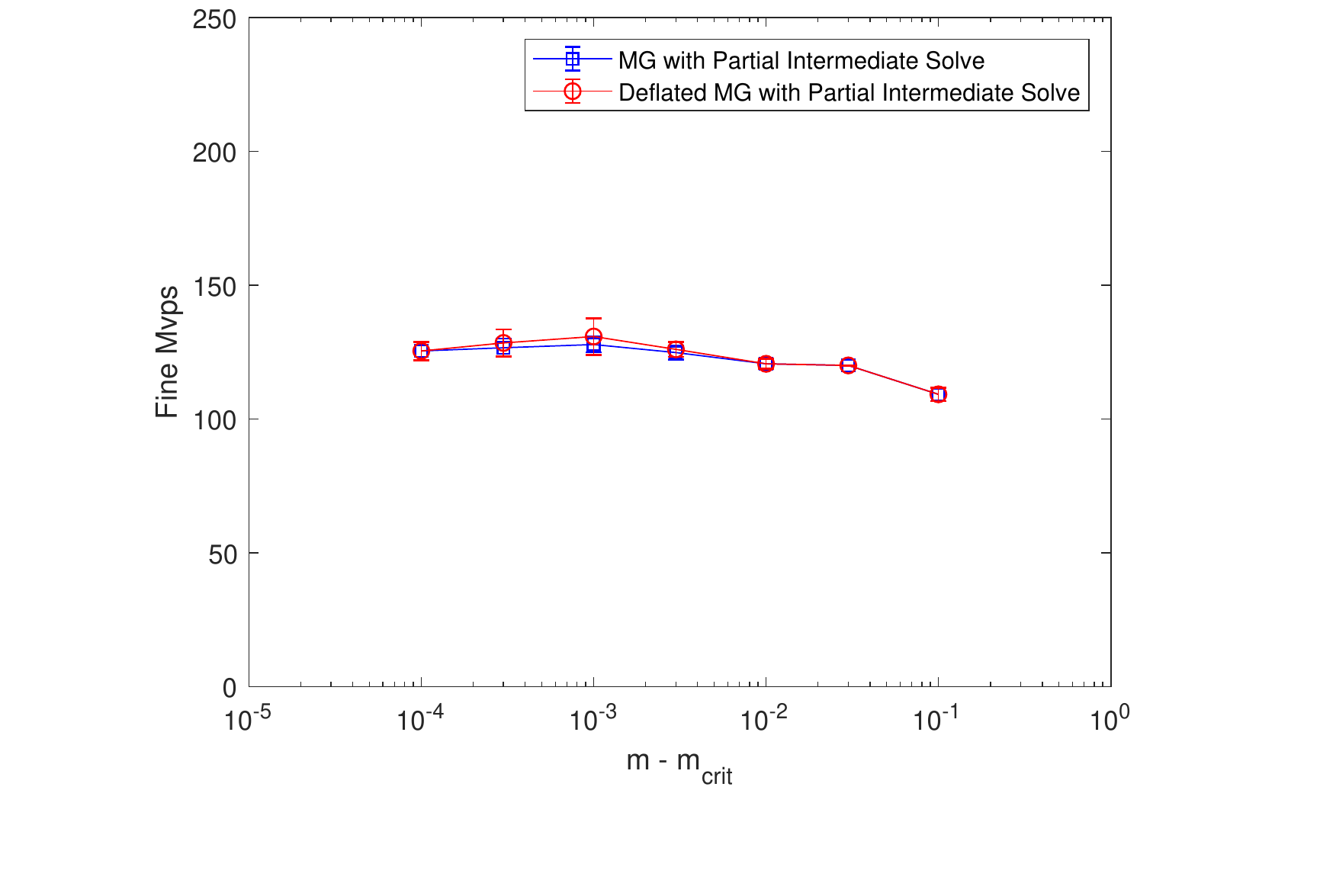}
\includegraphics[trim={2cm 8cm 2cm 9cm},clip,width=1.0\textwidth]{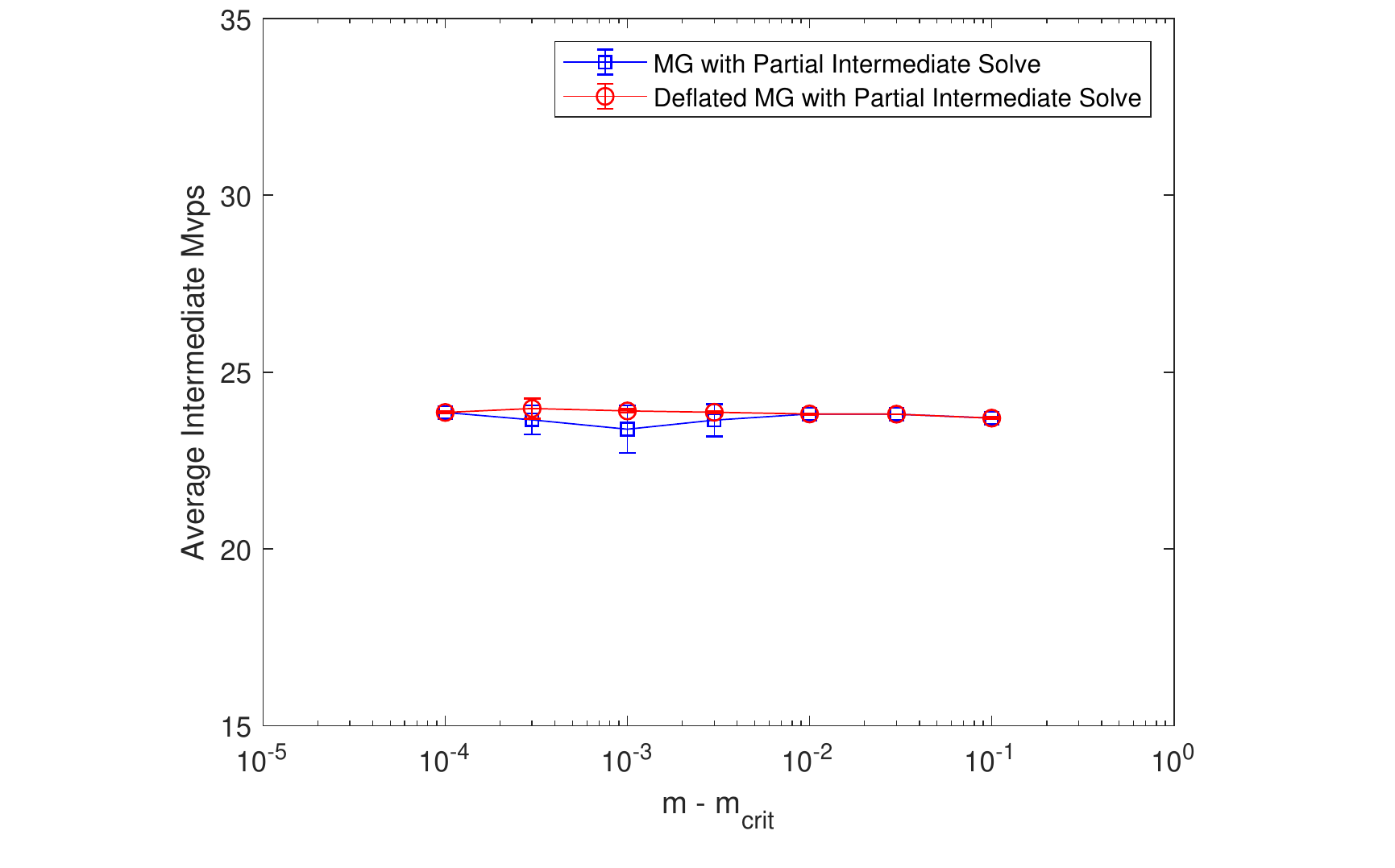}
\caption{(Top) The number of fine Dirac operator applications as a function of mass gap for both deflated and non deflated solves with the inclusion of a partial intermediate level solve for the lattice of size $256^2$. (Bottom) The average number of intermediate operator applications as a function of mass gap.}
\label{rad4}
\end{figure}
Figure 5 displays the fine equivalent Mvps as a function of lattice size at a mass gap of $10^{-4}$ on a log-log plot. The partial solves performed for our deflated method severely reduces the number of coarse operator applications, and subsequently lowers the number of fine equivalent Mvps. All methods display an approximate power behavior on lattice size. Our deflated MG method with partial solves exhibits the mildest dependence on lattice size. The estimate for the power in $\textrm{Mvps}=(\textrm{size})^{\alpha}$ is given in Table\ref{table1}.
 %space taken out; mvps to Mvps
\begin{figure}[ht]
\centering
\includegraphics[trim={2cm 8cm 2cm 8cm},clip,width=1.0\textwidth]{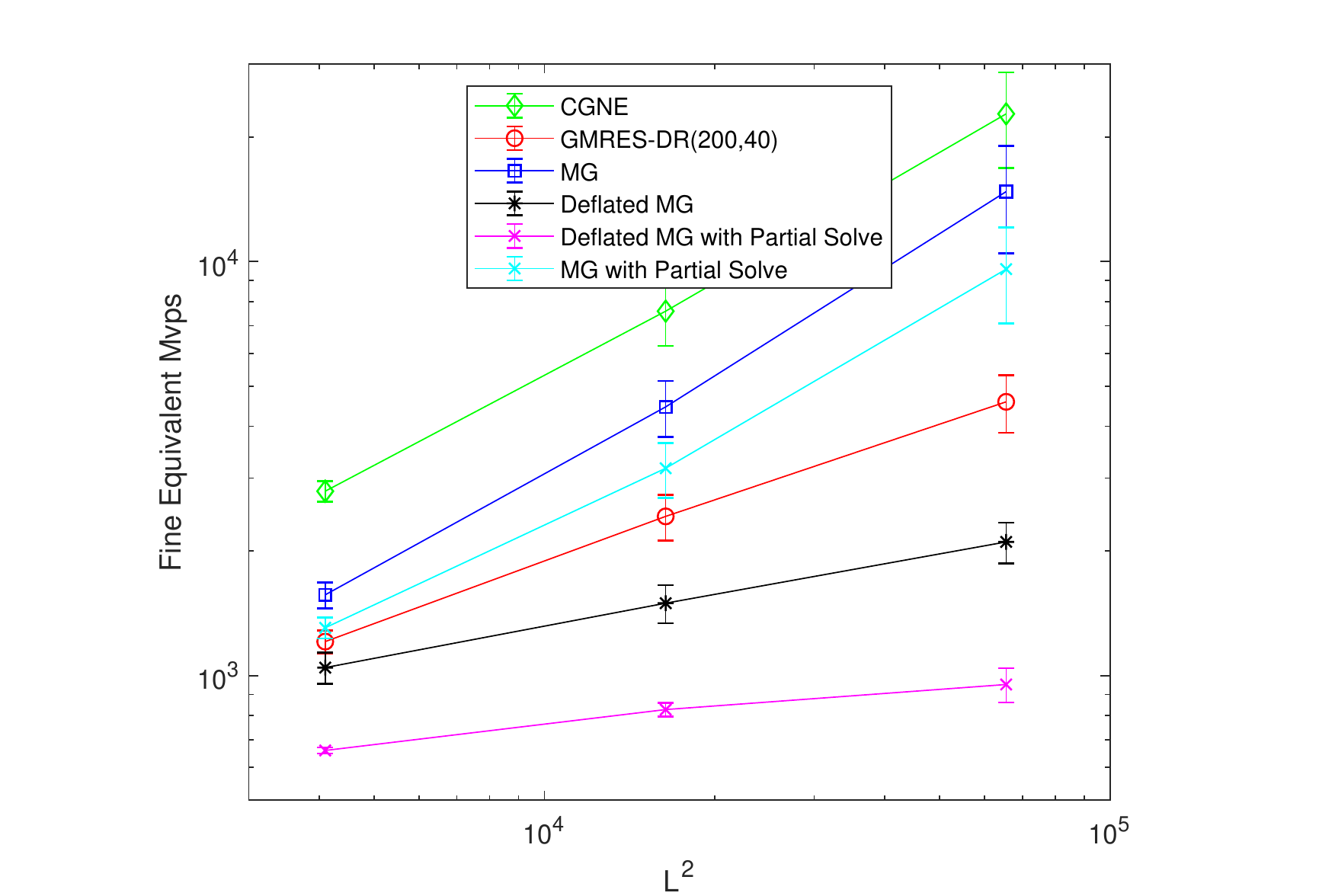}
\caption{Fine equivalent Mvps as a function of lattice size at a mass gap = $10^{-4}$.}
\label{rad5}
\end{figure}
% Figure 5 displays the fine equivalent Mvps as a function of lattice size at a mass gap of $10^{-4}$ on a log-log plot. The partial solves performed for our deflated method severly redcues the %number of coarse operator applications, and subsequently lowering the fine equivalent mvps. All methods display an approximate power behavior on lattice size. Our deflated MG method with %partial solves exhibits the mildest slope. The estimate for the power in $\textrm{Mvps}=(\textrm{size})^{\alpha}$ is given in Table\ref{table1}.

\begin{table}[!htpb]
    \centering
    \begin{tabular}{@{}l l  @{}}
        \toprule
          {$\textbf{Method}$}&  $\alpha$ \\ \midrule\midrule
        \textbf{DEFLATED MG WITH PARTIAL SOLVES} & $0.13 \pm 0.04$  \\
        \textbf{DEFLATED MG} & $0.25 \pm 0.05$ \\
        \textbf{GMRES-DR} & $0.48 \pm 0.06$  \\
        \textbf{NONDEFLATED MG WITH PARTIAL SOLVE} & $0.72 \pm 0.10$  \\
        \textbf{CGNE} & $0.76 \pm 0.10$  \\
        \textbf{NONDEFLATED MG} & $0.81 \pm 0.11$ \\ \bottomrule
    \end{tabular}
    \caption{Approximate powers in $\textrm{Mvps}=(\textrm{size})^{\alpha}$.}
    \label{table1}
\end{table}

\section{Conclusions}

Multigrid is an extremely effective algorithm to transfer critical slowing down to coarser operators, where it can be dealt with more efficiently. We have shown that the cost of a full solve on the coarse grid can be very large, but can be significantly reduced by a deflated and projected low tolerance solve. This method of deflation is more effective than deflation on the fine grid alone, without the increased storage costs associated with deflation. We also observe a characteristic synergy between MG and coarse grid deflation over multiple right hand sides that is not achieved by fine grid deflation or MG alone. Our method of deflation with partial solves shows a very mild dependence on lattice size, and is a significant step towards solving the strong scaling problem.

\section{Acknowledgements}

We thank the Baylor University Research Committee, the Baylor Graduate School, and the Texas Advanced SuperComputing Center for partial support. We thank the Wuhan conference organizers for student support as well as Jefferson National Laboratory for hosting Travis Whyte. We would also like to thank Paul Lashomb for his contribution to this work.

\section*{References}

\bibliography{referfile3}{}

\end{document}